\documentclass[a4paper,11pt]{article}

\pdfoutput=1

\usepackage{jcappub}
\usepackage{color}
\usepackage{latexsym}
\usepackage{amsfonts}
\usepackage{slashed}
\usepackage{array}
\usepackage{feynmp}
\usepackage{url}
\usepackage{subfigure}

\begin{document}
\title{IceCube, DeepCore, PINGU and  the indirect search for supersymmetric dark matter}

\author[a]{Paul Bergeron}
\author[b]{and Stefano Profumo}
\affiliation[a]{Department of Physics \& Astronomy, \\ University of Utah, 201 Presidents Circle, Salt Lake City, UT 84112, USA}
\affiliation[b]{Department of Physics and Santa Cruz Institute for Particle Physics,\\University of California, 1156 High St., Santa Cruz, CA 95064, USA}

\emailAdd{paul.bergeron@utah.edu}
\emailAdd{profumo@ucsc.edu}

\abstract{
\newline The discovery of a particle that could be the lightest CP-even Higgs of the minimal supersymmetric extension of the Standard Model (MSSM) and the lack of evidence so far for supersymmetry at the LHC have many profound implications, including for the phenomenology of supersymmetric dark matter. In this study, we re-evaluate and give an update on the prospects for detecting supersymmetric neutralinos with neutrino telescopes, focussing in particular on the IceCube/DeepCore Telescope as well as on its proposed extension, PINGU. Searches for high-energy neutrinos from the Sun with IceCube probe MSSM neutralino dark matter models with the correct Higgs mass in a significant way. This is especially the case for neutralino dark matter models producing hard neutrino spectra, across a wide range of masses, while PINGU is anticipated to improve the detector sensitivity especially for models in the low neutralino mass range.
}


\notoc

\maketitle

\section{Introduction}


These are exciting times for the exploration of the electro-weak scale. If the dark matter that dominates the matter budget in the universe is indeed a weak-scale weakly-interacting massive particle (WIMP), current experiments are closing in on its fundamental nature in multiple ways. First and foremost, the Large Hadron Collider (LHC) is very directly scrutinizing particles and interactions in the hundreds of GeV mass range, putting significant direct limits on theories, such as weak-scale supersymmetry, widely regarded as some of the theoretically best-motivated scenarios beyond the Standard Model of particle physics \cite{deJong:2012zt}. Additionally, the direct detection of WIMP dark matter is leapfrogging into the exploration of WIMP-nucleon cross sections that are also predicted by theories such as supersymmetry. This progress is placing unprecedented pressure on certain corners of the parameter space of theories such as the minimal supersymmetric extensions of the Standard Model (MSSM) \cite{Baudis:2012ig}. While indirect dark matter detection (for a pedagogical introduction see e.g. \cite{Profumo:2013yn}) is more prone to astrophysical uncertainties than other search methods, indirect searches represent an increasingly promising avenue to explore the nature of dark matter and, more broadly, of the electroweak scale.

The recent discovery of a boson that matches the properties expected from the Standard Model Higgs (or in certain limits, of one of the MSSM Higgses) has not only added an additional piece to the puzzle of what lies beyond the Standard Model, but also knowledge of which corner of the MSSM might be realized in nature, if any \cite{cms, atlas}. The mechanism of electroweak symmetry breaking is also being directly tested, with profound implications for both particle physics and cosmology.

The MSSM is currently undergoing close scrutiny both from direct searches for supersymmetric particles, and from the standpoint of the ability of the theory to produce a Higgs with properties compatible with observation \cite{cms, atlas}. In addition, MSSM neutralino dark matter is under the additional pressure of results from direct detection, especially from the Xenon100 experiment \cite{Aprile:2012nq}, which rules out portions of the MSSM parameter space where the correct thermal relic abundance is obtained by a finely tuned mixing of the bino and higgsino components of the lightest neutralino.

Among indirect WIMP dark matter detection experiments, perhaps the smallest astrophysical uncertainties affect the search for high-energy neutrinos originating from the annihilation of WIMPs captured inside the Sun or the Earth. In the limit of equilibrated capture and annihilation, the flux of neutrinos only depends on the rate at which WIMPs are captured; in the case of the Sun, the WIMP capture rate depends, in turn, primarily on the spin-dependent WIMP-nucleon cross section (this is true for many particle dark matter models, including MSSM neutralinos). With the advent of km$^3$-size neutrino telescopes, this search channel has started to constrain directly many WIMP dark matter models, contributing to our understanding of which type of particle could actually play the role of the dark matter.

Of recent results from neutrino telescopes, the most impressive limits on particle dark matter come from the IceCube telescope, supplemented with the DeepCore subarray \cite{abbasietal}. We will comment and detail on those below, but suffice it to say that these results carve into important regions of the MSSM neutralino parameter space.

In addition, the IceCube array will soon be complemented with an even more thickly instrumented section than the DeepCore subarray, called PINGU, that promises to further lower the energy threshold for the detection of muons produced by charged-current interactions of high-energy neutrinos. With this additional subarray, the potential for dark matter searches might be further enhanced.

One of the questions we intend to address here is how this plays out when factoring in the recent direct limits on supersymmetric particles and the discovery of the Higgs boson. In principle, the measurement of the Higgs mass can directly affect the flux of neutrinos from the Earth or from the Sun, as it impacts the mass scale of scalars. For example, if the neutralino is bino-like, scalar masses impact the relevant scattering cross sections directly. Secondly, the Higgs mass measurement constraints, again for bino-like neutralinos, the mass range associated with a resonant light, CP-even Higgs s-channel exchange as the main process setting the thermal relic density in the early universe. 

The present study is motivated by the following two goals:

(i) to re-evaluate the role of neutrino telescopes in the search for supersymmetric neutralino dark matter in light of the LHC results, and, specifically, in connection with direct particle searches, and

(ii) to assess the role of PINGU as a tool to search for neutralino dark matter, and to single out in which corners of the MSSM parameter space  will PINGU offer the greatest benefits and be most critical.

The remainder of this paper is structured as follows: in Section~\ref{sec:telescopes} we introduce the key experimental apparata, namely the IceCube/DeepCore/PINGU telescope; in Section~\ref{sec:scan} we describe the procedure we employ to explore the MSSM parameter space, and the limits we impose on each supersymmetric model; Section~\ref{sec:results} describes our results, and, finally, Section~\ref{sec:conclusions} contains our discussion and conclusions.

\section{The IceCube/DeepCore/PINGU Telescope}\label{sec:telescopes}
IceCube is a second generation, kilometer-cubbed-scale, neutrino telescope situated in the Antarctic. The telescope itself relies on detection of Cherenkov radiation from muons created by neutrino interactions with sterile antarctic ice~\cite{halzenhooper}. This telescope consists of a total of 86 vertical strings each with Digital Optical Modules (DOM's) that contain a digitizer board and a photomultiplier tube (PMT). 78 of these strings carry 60 DOM's, each placed at 17~m intervals from a depth of 1450~m to 2450~m below the ice surface~\cite{abbasietal}.

In addition to the main 78 strings for IceCube, the other 8 are infill-- specialized for a subarray dubbed DeepCore~\cite{Brown:2010cs}. DeepCore has been deployed in primarily the bottom half of the central region of IceCube~\cite{halzenhooper}, and completed its first year of data collection in 2011~\cite{DeYoung:2011ke}. 6 of the 8 infill strings each have 50 specialized DOM's spaced in intervals of 7~m. These specialized DOM's are a higher quantum efficiency version Hamamatsu PMT (HQE DOM), differing from the rest of the PMT's by using a ``super bialkali'' photocathode. The HQE DOM improves the quantum efficiency of the original PMT by about 40\% for a photon with a wavelength of 390 nm (based on Hammatsu laboratory tests)~\cite{Collaboration:2011ym}, and 35\% when averaged over the entire spectrum of detected Cherenkov light (based on \emph{in situ} measurements). The other two infill strings have a mixture of the regular IceCube DOM and the newer DeepCore HQE DOM. The total number of strings in DeepCore is 20: the 8 infill strings together with 12 strings pre-existing from IceCube~\cite{Brown:2010cs}. Aside from the increased DOM linear density in DeepCore, the strings themselves are packed much more densely in space than they are for IceCube. 13 of the 20 strings have an average horizontal inter-string distance of 72~m, while 6 of the 20 have an average of 42~m inter-string distance~\cite{DeYoung:2011ke}~\cite{Collaboration:2011ym}. This compares to IceCube, whose strings have an inter-string distance of 125~m~\cite{abbasietal}. 

\emph{In situ} measurements using pulsed LED sources were used to determine the placement of DeepCore at 2100 to 2450 meters below the surface of the polar icecap. This location was selected due to increased transparency of the ice ($\sim$~40\% - 50\% increase) and to avoid a dust layer. The dust layer, resulting from an interstadial period in the last glacial period, exists from 2000 to 2100 meters below the surface. Consequently, DeepCore avoids this high absorption and scattering area. The remaining 10 DeepCore DOM's are deployed on the 8 specialized strings at 10 m intervals above the dust layer~\cite{Collaboration:2011ym}.

From the success of IceCube and DeepCore seen so far, a further extension to the detector is now under consideration. This extension, the Precision IceCube Next Generation Upgrade (PINGU), will additionally increase the PMT density in a region within DeepCore. Current plans for PINGU, whose final detailed proposed design is not public yet (to our knowledge), would add roughly 20 more strings similar to those already employed by DeepCore. Some strings may incorporate a new set of PMTs, which are currently being developed for the proposed KM3NeT detector, instead of those in the HQE DOM's~\cite{DeYoung:2011ke}.

In addition to the DOM's, the IceCube detector also employs IceTop: an array of 160 ice filled, Auger-style, Cherenkov detectors. These help  identify neutrinos due to cosmic ray events, and they are employed as well for calibration and cosmic ray studies~\cite{halzenhooper}. However, IceCube can also make the distinction between solar neutrinos and cosmic ray events by simply selecting upwards-going muon tracks while the Sun is below the horizon~\cite{abbasietal}. 

The main array of IceCube, while focused on TeV energy neutrinos and above~\cite{DeYoung:2011ke}, is expected to be sensitive to neutrinos down to energies of 100 GeV~\cite{halzenhooper}; DeepCore, meanwhile, is sensitive to neutrino energies as low as 10 GeV~\cite{DeYoung:2011ke}~\cite{Collaboration:2011ym}. This allows the IceCube detector to be effectively sensitive to neutralinos down to masses of about 50 GeV~\cite{halzenhooper}~\cite{abbasietal}. If PINGU is built, the additional DOM density will lend IceCube the ability to detect neutralinos with masses below 30 GeV and potentially as low as a few GeV~\cite{DeYoung:2011ke}. In comparison, the ANTARES telescope, a kilometer scale neutrino telescope sunk in the mediterranean, has a threshold on the order of 10 GeV~\cite{Bertin:2011zz}.

Among other science goals, IceCube is actively searching for neutrino signals of dark matter annihilations. Instead of relying on areas of the dark matter halo with expected neutralino densities significant enough for indirect detection, IceCube will look at the Sun.  Despite the fact that WIMPs interact extremely weakly, occasional scattering of a particle off of the Sun's matter can cause enough of a reduction in the particle's velocity so that the scattered WIMP may become gravitationally bound inside the Sun. Such WIMPs will then accumulate over time in the center of the Sun, eventually possibly reaching equilibrium between capture and annihilation~\cite{halzenhooper}~\cite{abbasietal}, depending upon the particle physics properties of the WIMP.

Over its lifetime, the Sun has moved about the galactic halo a multitude of revolutions. Because of this motion, any asymmetries within the galactic halo will have been averaged out, and therefore halo structure is not of concern. This also means that IceCube and DeepCore will not rely on looking at halo regions of high density, where distance reduces incident flux and events could be due to unknown physics (such as phenomena at the galactic center), as is relied on by indirect searches with, e.g., gamma rays~\cite{halzenhooper}.

Furthermore, IceCube offers the ability to effectively constrain a different set of models than those of interest to the major direct detection experiments, in particular experiments primarily sensitive to spin-independent WIMP-nucleon interactions. As mentioned above, IceCube will be most sensitive to spin-dependent scattering as this typically dominates the capture process in the Sun~\cite{halzenhooper}~\cite{abbasietal}~\cite{DeYoung:2011ke}. This complementarity between spin-independent direct searches and searches with neutrino telescopes will continue to hold in the future: even with next-generation direct detection experiments, and even restricting one's attention to supersymmetric models only, one would still miss detection of models that could have detection rates of $\sim$1000 events per year at IceCube~\cite{halzenhooper}. The privileged access to models with large spin-dependent WIMP-nucleon cross section therefore makes the IceCube observatory a unique tool among dark matter detection experiments.

\section{Scan of Supersymmetric Parameter Space}\label{sec:scan}
%
%
We perform a scan over the low-energy (i.e. weak-scale) parameter space of the minimal supersymmetric extension to the Standard Model (MSSM). We use the $\mathtt{DarkSUSY}$ package~\cite{Gondolo:2004sc}, and we generate models randomly based on the subroutine \emph{random}\_\emph{model}. For this scan, we employ the so-called ``phenomenological'' MSSM, whose entries are defined by the parameters specified in Table~\ref{tab:parameters}, where the masses have all units of GeV. For the soft supersymmetry-breaking gaugino masses, we allowed the input masses to take on opposite signs. Triscalar couplings are given in units of the universal scalar soft-supersymmetry breaking masses. The first and second generation triscalar couplings are universally set to zero. We performed the scan logarithmically over the range specified in Table~\ref{tab:parameters} for all parameters with the exception of $A_t$ and $A_b$ which were scanned linearly.

\begin{table}[b]
  \centering
  \begin{tabular}{| l c | c |}
    \hline
      Parameter                                                              &                            & Range \\ \hline
      CP-odd Higgs                                                       & $m_A$                & $\pm(250,\ 50000)$\\
      $\mu$                                                                     & $\mu$                & $\pm(217,\ 10000)$\\
      Bino and Wino Masses                                       & $M_1$,$M_2$ & $\pm(217,\ 20000)$ \\
      Gluino Mass                                                          & $M_3$              & $\pm(500,\ 20000)$ \\
      $\tan(\beta)$                                                           & $\tan(\beta)$    & $(2,\ 60)$ \\
      $1^{st}$ and $2^{nd}$ generation sfermions & $m_{\widetilde {q}_{1,2}}$              & $(217,\ 20000)$ \\
      $3^{rd}$ generation sfermions & $m_{\widetilde q_3}$              & $(217,\ 20000)$ \\
      $3^{rd}$ generation triscalar squark  coupling                       & $A_t$, $A_b$               & $(-2.0,\ 2.0)$ \\
    \hline
  \end{tabular}
  \caption{MSSM parameters that were varied randomly in computing the parameter space scan, within the ranges indicated in the right-hand column. All entries are given in units of GeV, with the exceptions of $\tan(\beta)$ (unitless), and $A_t$ and $A_b$ (units of $m_{\widetilde {q_3}}$).}
  \label{tab:parameters}
\end{table}

Each model is tested against the default requirements of having the lightest neutralino as the lightest supersymmetric particle, and of passing the standard $\mathtt{DarkSUSY}$ constraints for precision measurement, as well as collider constraints~\cite{Hagiwara:2002fs, Beringer:1900zz}.  The constraints we implement on our parameter space scan via {\tt DarkSUSY} include: direct collider searches for charginos, gluinos, squarks and sleptons with LEP-2, the invisible $Z$ width, the $\rho$ parameter, $(g-2)_\mu$, and $b\to s\gamma$. We also require the Higgs mass to be within the 2$\sigma$ range ($123$~GeV~$< m_h < 128$~GeV) identified by \cite{atlas, cms}. Additional searches for supersymmetric particles at the LHC are implemented by conservatively requiring gluinos and squarks to be generically heavier than 1.5 TeV, in accordance with recent results \cite{ATLAS:2013lla}. In addition, in all plots we implement on a model-by-model basis, the constraints from the XENON100 results \cite{Aprile:2012nq}.

We enforce an {\em upper} limit to the lightest neutralino thermal relic density, which we set at $\Omega_\chi h^2=0.135$. A lower bound is not applied as models with an under abundant relic density can still be provide the entire dark matter by, for example, non-thermal production~\cite{Profumo:2004at}. Note that we thus assume that for all models neutralinos make up 100\% of the dark matter in the universe. Wherever relevant, we also duplicate our results focusing exclusively on those models that have a {\em thermal} relic density of neutralinos in agreement (at WMAP7's 2$\sigma$ level~\cite{Jarosik:2010iu}; $0.1053<\Omega_\chi h^2<0.1165$) with the observed universal dark matter density. Although a non-standard cosmological thermal history could make under- and over-abundant models viable, we hold the thermal production mechanism as especially strongly motivated, and thus worth highlighting. We additionally impose an upper bound on the neutralino mass set at 2 TeV. This choice was made to avoid regions of the parameter space where non-perturbative electroweak effects, which are not included in the $\mathtt{DarkSUSY}$ package, produce significant  contributions to the pair-annihilation of neutralinos~\cite{Profumo:2005xd}. This was seen especially for wino-like models, but we uniformly applied the upper mass limit to all neutralino types.

As far as the density and velocity distribution of Galactic dark matter is concerned, we adopt here the default choices of 
$\mathtt{DarkSUSY}$ (a Navarro-Frenk-White density profile with a Maxwellian velocity distribution).

We classify each ``viable'' model according to the lightest neutralino $\chi_1^0$ mixing fraction in terms of its bino, wino and higgsino fraction. If the mass eigenstate $\chi_1^0$ is given in terms of interaction eigenstates by the linear combination
\begin{equation}
\chi_1^0=N_{11}\widetilde B+N_{12}\widetilde W^0+N_{13}\widetilde H^0_1+N_{14}\widetilde H^0_2
\end{equation}
then the ``bino fraction'' is defined as $|N_{11}|^2$, the ``wino fraction'' as  $|N_{12}|^2$, and the ``higgsino fraction'' as $1-|N_{11}|^2-|N_{12}|^2$.  If a neutralino composition fraction is greater than 91\%, then the corresponding model is classified as a pure neutralino of a type that corresponds to the largest mixing ratio. If instead it is less than 91\%, we define the neutralino to be a mixed state of the two types which have the greatest contribution to the neutralino state. For example a neutralino with a bino fraction of 92\% is classified as a \emph{bino-like} neutralino; one with a fraction of 63\% wino-like, 18\% bino-like, 19\% higgsino-like would be classified as a \emph{wino/higgsino-like} neutralino. 

To have a balanced sample of neutralinos of all compositions, we biased our scans to obtain more of one or the other neutralino types (we are not interested here in any statistical statement about the MSSM parameter space). For example, to obtain more wino-like neutralinos, the parameter $M_{2}$, with a range 1.5 to 3.0 TeV, was set as the lower bound for $\mu$, $M_{1}$, $M_{3}$, and $m_{\tilde q}$. Similarly, for the higgsino-like scan, we sampled $\mu$ within the range of 0.7 to 1.5 TeV, and set it as the lower bound for $M_{1}$, $M_{2}$, $M_{3}$, and $m_{\tilde q}$. 

\begin{figure*}[t]
  \centering
    \includegraphics[scale=0.9]{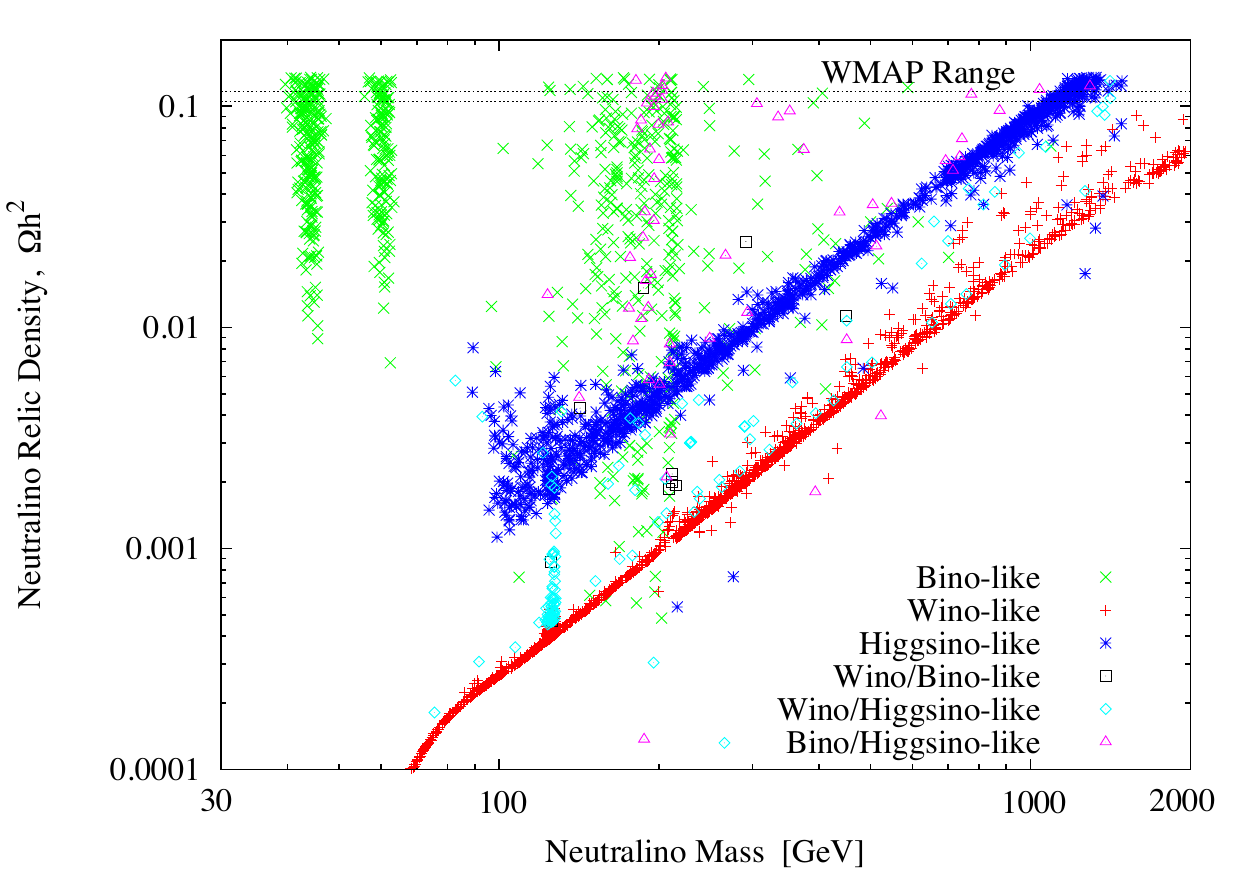}
    \caption{The results of the parameter space scan in terms of neutralino relic density versus lightest neutralino mass, broken down by different neutralino compositions, color-coded as in the legend.}
    \label{fig:DensityvMass}
\end{figure*}

We show the results of our scan, in terms of the neutralino relic density versus neutralino mass, in Figure~\ref{fig:DensityvMass}; different neutralino compositions are shown with different symbols. As expected from previous studies (e.g. Ref.~\cite{Baer:2005jq}), we find that there is a sharp transition from bino to wino-like dominance in the parameter space in regards to $M_1$ and $M_{2}$. As $M_{1}$ becomes large with respect to $M_{2}$,  the dominant neutralino type quickly transitions from bino-like models to wino-like models, with very few mixed bino-wino models. We also do not find any bino-like neutralinos around 50 GeV and in the range 70 to 100 GeV, due to relic density constraints and the lack of coannihilating partners forced by direct collider searches for supersymmetric particles in the relevant mass range. We do find two bino-like regions in neutralino mass, corresponding to resonant $s$-channel annihilations with the Z boson and Higgs boson, respectively, as $2m_{\chi}\approx~M_{Z}$ and $2m_{\chi}\approx~m_{h}$ respectively~\cite{Baer:2005jq}.

It is also clear from the scan, as depicted in Figure~\ref{fig:DensityvMass}, that MSSM neutralinos with masses below 100 GeV predominantly exist in highly bino-like compositions. Above 200 GeV, and again due to the requirement of a sufficiently low thermal relic density, bino-like compositions become increasingly less likely with increasing mass, with heavier neutralinos primarily appearing as mixed states. Bino mixed dark matter also has the greatest mass extent consistent with the WMAP range (shown by the horizontal black dotted lines in the firgure) for the dark matter relic density; higgsino, wino, and wino-bino states appear within the WMAP range only for masses at or above a TeV.  

\section{Results}\label{sec:results}

\begin{figure*}[t]
  \centering
  \mbox{
      \subfigure[]{
          \includegraphics[scale=0.6]{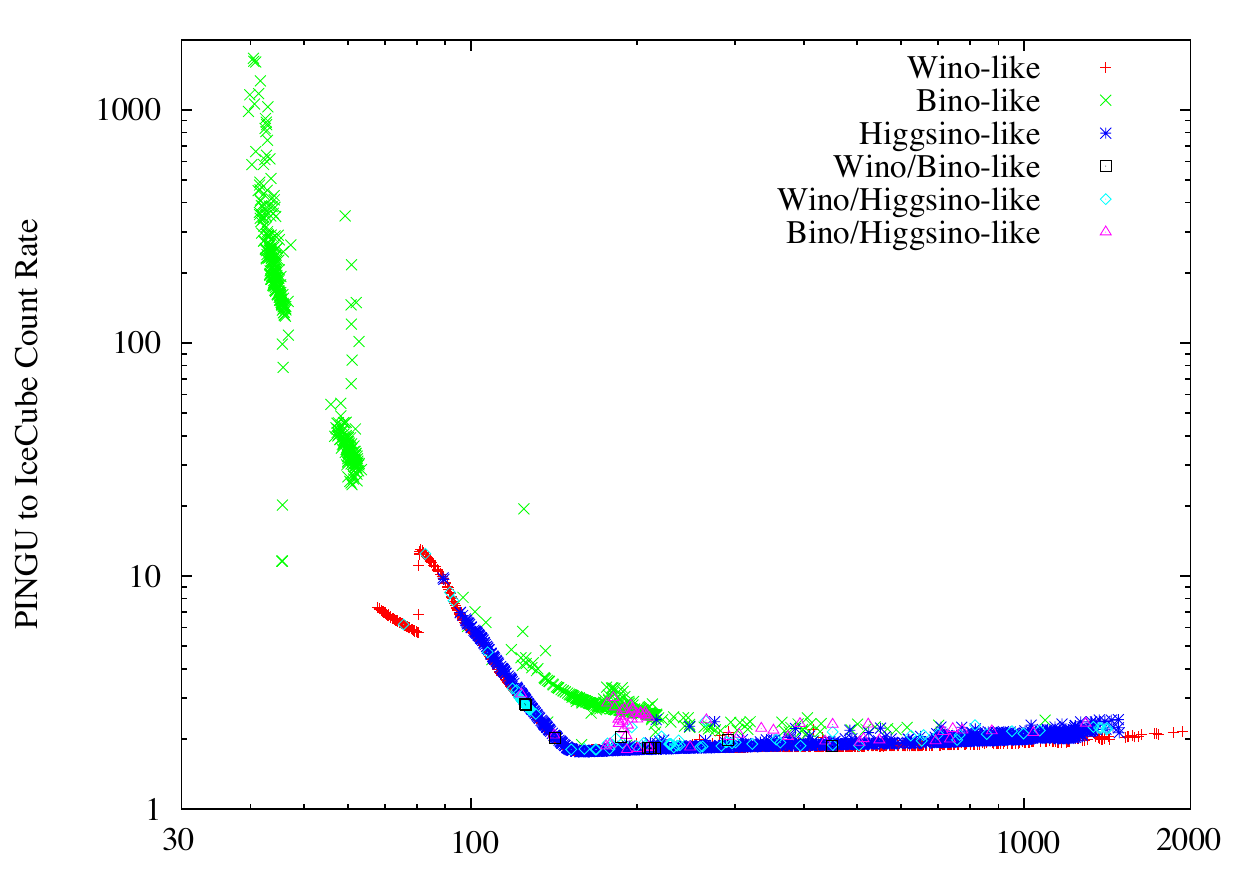}
      }
      \subfigure[]{
        \includegraphics[scale=0.6]{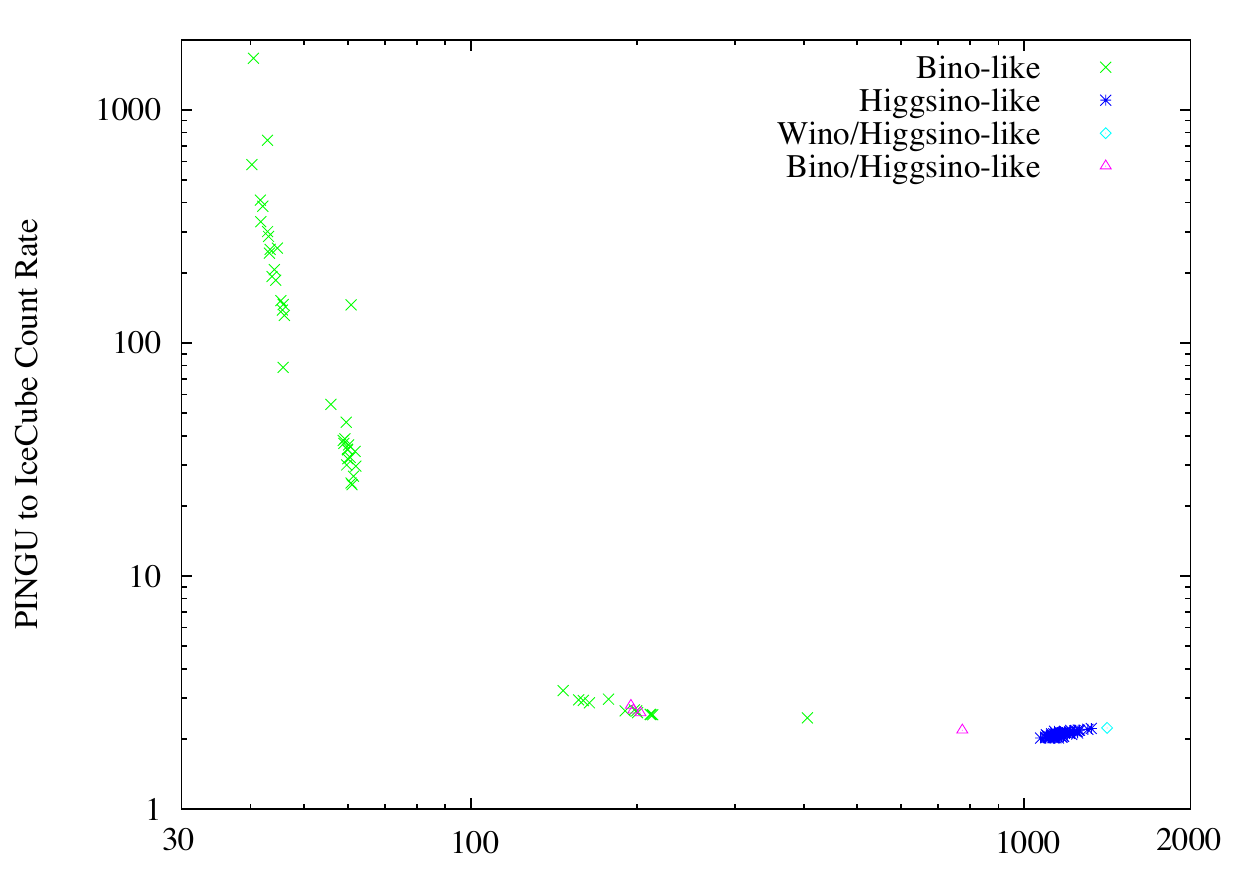}
      } 
    } \\ 
    \mbox{
      \subfigure[]{
          \includegraphics[scale=0.6]{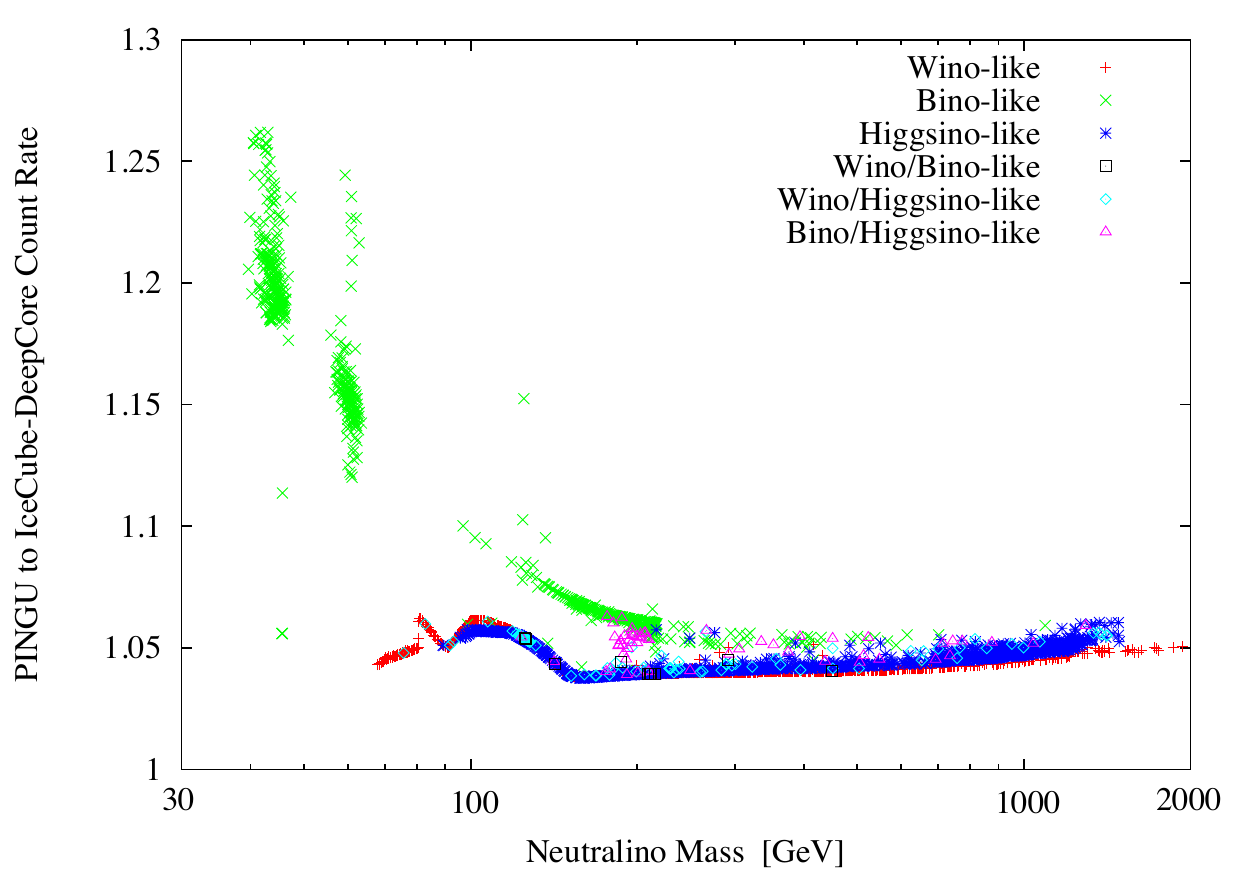}
      }
      \subfigure[]{
        \includegraphics[scale=0.6]{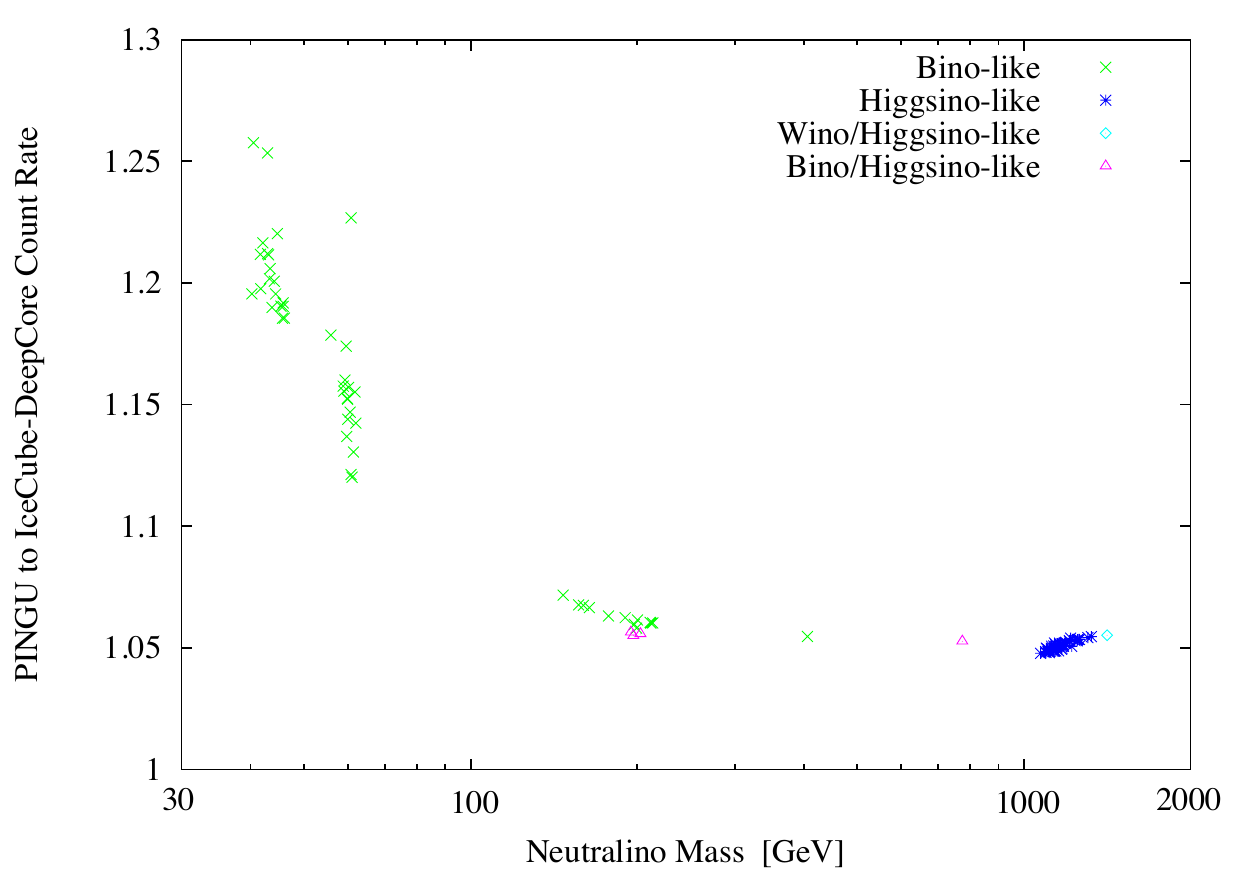}
      } 
    } 
  \caption{Plotted is the ratio of the neutrino count rate of the IceCube Telescope with PINGU to the IceCube telescope without PINGU for the various MSSM neutralino models shown in Figure~\ref{fig:DensityvMass} (\emph{Panels (a) and (c)}) as well as only those in agreement with the WMAP relic density limits (\emph{Panels (b) and (d)}), with the same color- and symbol-coding as Figure~\ref{fig:DensityvMass}. \emph{Panels (a) and (b)}: The configuration here is for IceCube {\em without} the DeepCore subarray. \emph{Lower Panel}: The configuration here is now for IceCube {\em with} DeepCore.}
  \label{fig:flux}
\end{figure*}

The upgrade of the IceCube detector with PINGU will increase the detector sensitivity to lower mass dark matter models by increasing the effective volume of IceCube at low energy through increased, albeit localized, string density. Figure~\ref{fig:flux} compares the IceCube neutrino count rate (for muon energies above 1 GeV) with and without PINGU. The upper panel refers to IceCube only strings, whose effective area we take as equivalent to $V_{eff}^{2/3}$, with the effective volume as given in~\cite{Wasserman}. The lower panel makes use of IceCube {\em including} the DeepCore Subarray, whose effective area is given in~\cite{Collaboration:2011ym}. The effective area for IceCube with PINGU was instead taken from Ref.~\cite{Wasserman} and is used for all panels. We consider separately the sensitivity improvement that PINGU will bring to IceCube and DeepCore, since each of the three are separate detectors from one another. The question we are thus asking is how PINGU by itself would perform in addition to IceCube or in addition to IceCube plus DeepCore. The two panels to the right indicate the same results, this time exclusively for models with a thermal neutralino relic density consistent, to the 2$\sigma$ level, with the WMAP 7 year results~\cite{Jarosik:2010iu}. 

Our results clearly show the substantial increase in the sensitivity of IceCube to a neutrino flux from dark matter annihilation that is brought about by lowering the energy threshold needed for neutrino detection. Thanks to PINGU alone (upper panel of Figure~\ref{fig:flux}) the performance of IceCube is boosted to a relative count rate increase of 20 to 1000 times for masses below 70~GeV. 

The addition of PINGU to the IceCube+DeepCore array, which we show in the lower panel, is less dramatic, but still significant for certain MSSM neutralino models. With PINGU, the flux increase is expected to be  of about 5\% for most models (originating from the very low energy neutrinos), while providing a significant increase for bino-like neutralinos with masses below 70 GeV, by as much as $\sim$30\%.  

We note that had we relaxed the assumption of slepton-squark universality in our scan, our results might have been significantly affected. In fact, unlike squarks, sleptons are only mildly constrained by LHC results \cite{Dutta:2013sta}, and if universality is dropped their mass can be as low as permitted by direct LEP searches for slepton pair production (in the vicinity of 100 GeV). As a result, for example, slepton coannihilation could render light bino-like neutralinos viable dark matter candidates. Such light bino-like states are precisely the type of dark matter candidates where PINGU would play an ever more important role, which is here blurred by our squark-slepton universality assumption.

\begin{figure*}
  \begin{tabular}{c | c}
    \includegraphics[scale=0.6]{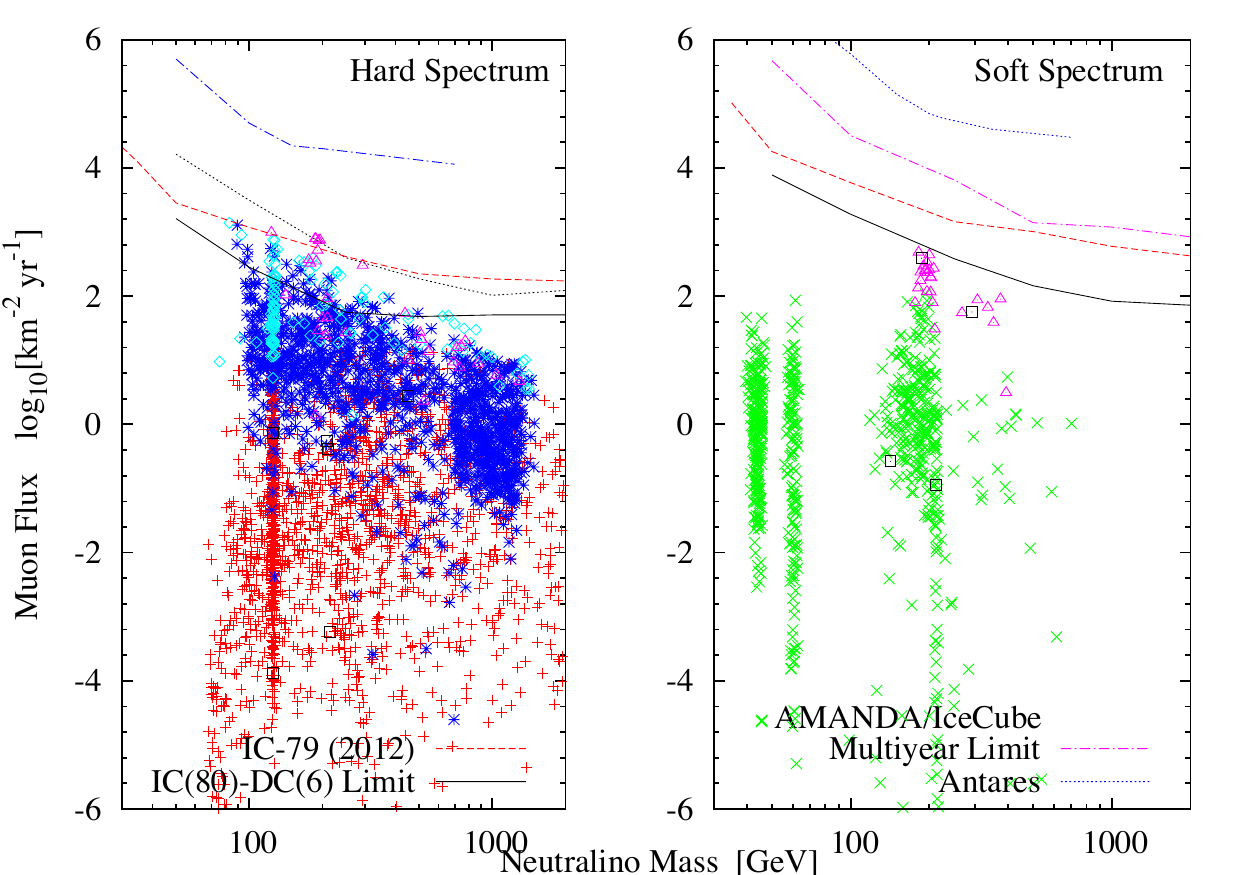} & \includegraphics[scale=0.6]{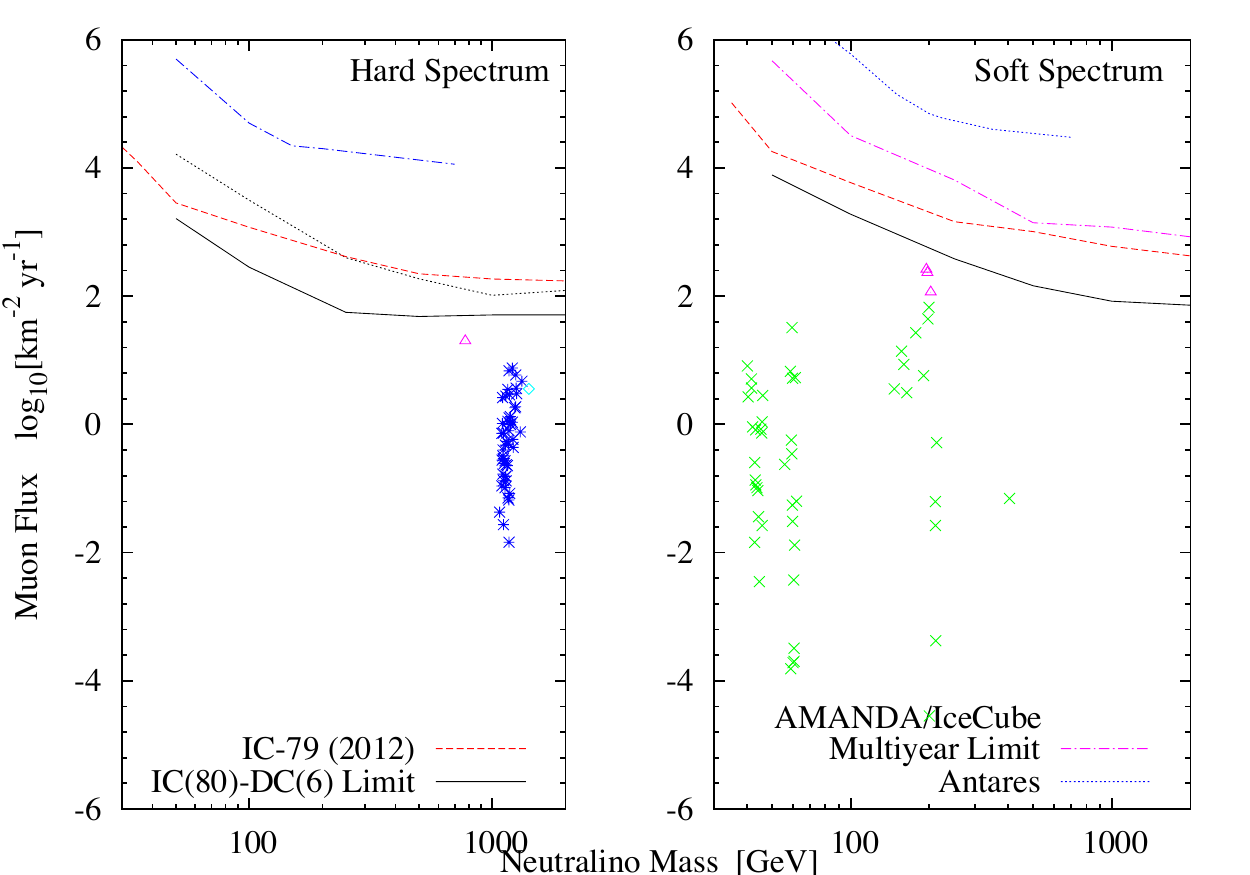}
  \end{tabular}
  \caption{Here we show the results of the scan with the muon neutrino flux as a function of neutralino mass for decays via hard (\emph{panels 1 and 3}) and soft (\emph{panels 2 and 4}) channel decays. The \emph{left two panels} use all models from Figure~\ref{fig:DensityvMass} while the \emph{right two panels} are only those models with the correct thermal relic abundance. The models are classified based on neutralino composition, with symbols and colors corresponding to the same scheme used in Figures~\ref{fig:DensityvMass}~and~\ref{fig:flux}. Curves are shown for limits set by searches with ANTARES~\cite{Mangano:2012mc}, AMANDA-IceCube (multiyear limit~\cite{IceCube:2011aj}), IceCube plus Deep Core, in the 86-String Configuration~\cite{Danninger}, and IceCube-79 (2012)~\cite{:2012ef}.}
  \label{fig:muon}
\end{figure*}

In Figure~\ref{fig:muon} we present the results of the muon flux rates from the scan, and compare against current neutrino telescope performances. We use the same color coding as in Figure~\ref{fig:flux} to distinguish among the various neutralino composition, and, again, the two panels to the right refer to models with the right thermal relic density. Comparing with limits from ANTARES~\cite{Mangano:2012mc}, AMANDA and IceCube~\cite{IceCube:2011aj}, and IceCube~\cite{Danninger}~\cite{:2012ef}, we see that the IceCube/DeepCore system is beginning to significantly cut into the parameter space of the MSSM, for models with the right Higgs mass and a low-enough relic density. The figure distinguishes between a ``hard'' (\emph{left}) and a ``soft'' (\emph{right}) neutrino spectrum, the former associated with higgsino or wino (or mixed states containing a significant higgsino or wino fraction) while the latter associated with predominantly bino-like neutralinos. Even for a soft spectrum, MSSM neutralinos are starting to fall within the reach of the IceCube/DeepCore system. In particular, our results indicate that a significant fraction of neutralinos featuring a significant Higgsino fraction (e.g. purple triangles, indicating bino-higgsino neutralinos, blue stars, indicating pure higgsinos, and light blue diamonds, indicating wino-higgsino mixed states) are currently either ruled out by IceCube+DeepCore, or on the verge of being ruled out. MSSM neutralino models are accessed and constrained across a wide range of masses (unlike other indirect detection methods, where low masses are usually best constrained) from 80 GeV to over a TeV in neutralino mass.

\begin{figure*}
    \centering
    \includegraphics[scale=0.6]{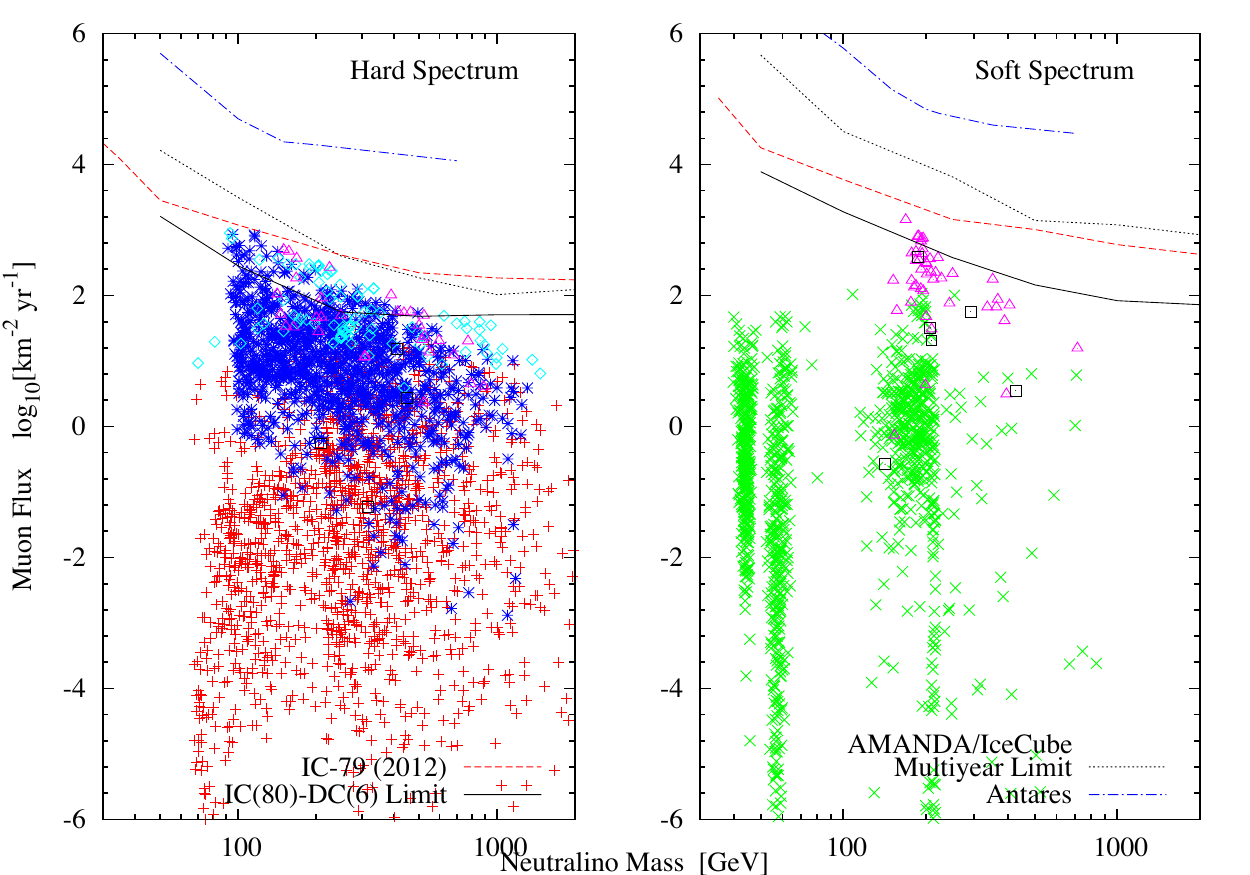}   
    \caption{As in Figure~\ref{fig:muon}, but for a broader scan that includes models with a lightest CP-even Higgs whose mass is in accordance with the LEP bound of $>114$~GeV.}
  \label{fig:nohiggs}
\end{figure*}

In Figure~\ref{fig:nohiggs} we duplicate what shown in Figure~\ref{fig:muon}, this time for a much broader scan which includes models with a lightest CP-even Higgs both within and outside the LHC Higgs mass measurement range (i.e. in accordance with the lower bound of 114~GeV set by the LEP). The figure therefore visually shows the impact of the Higgs discovery on the flux of neutrinos from the Sun within supersymmetric dark matter models. We find that the rates and general qualitative features of our findings are unchanged, with the only differences being: (i) a broader scatter around the Higgs resonance for bino-like neutralinos, and (ii) a comparatively larger fraction of models featuring a larger muon flux (as expected from a more generous scalar mass range). 

\begin{figure*}
  \begin{tabular}{c | c}
    \includegraphics[scale=0.6]{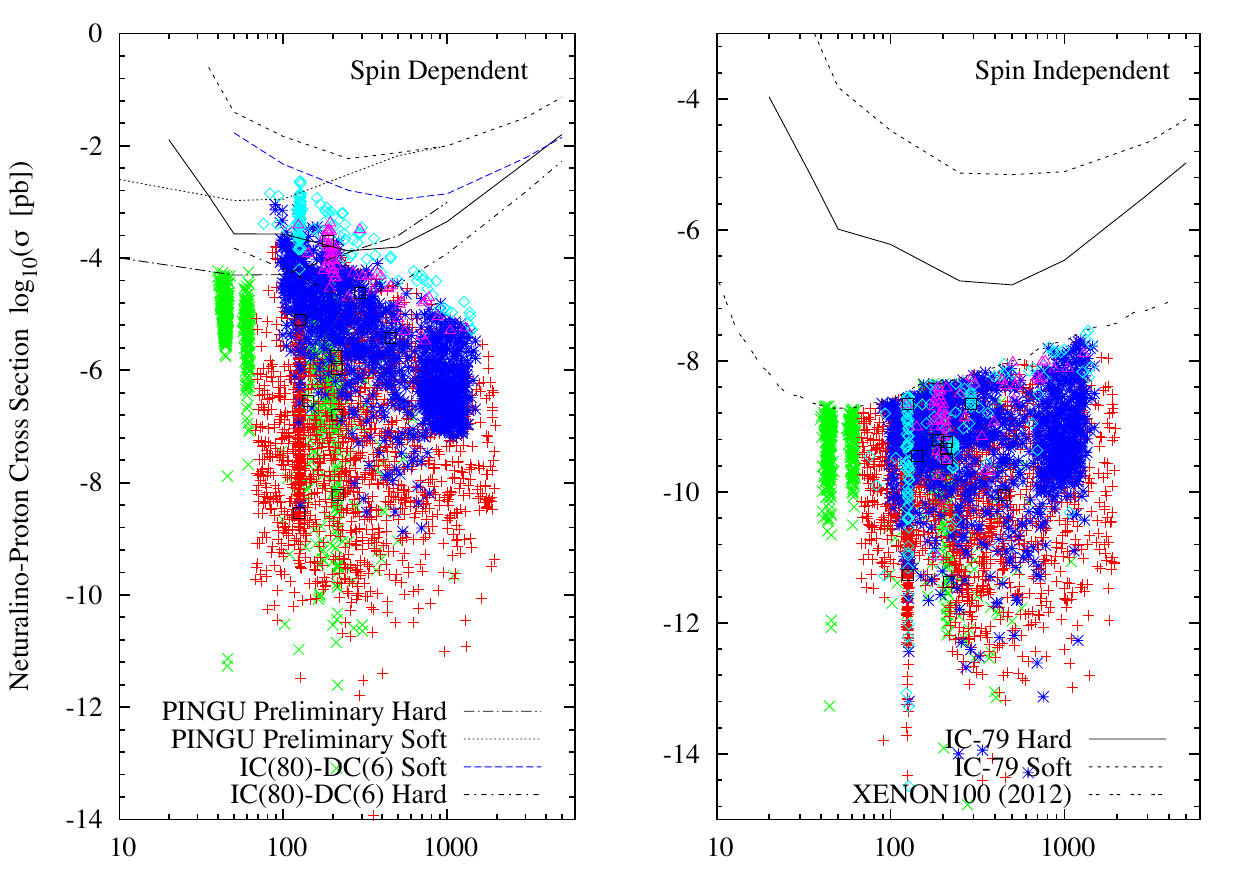} & \includegraphics[scale=0.6]{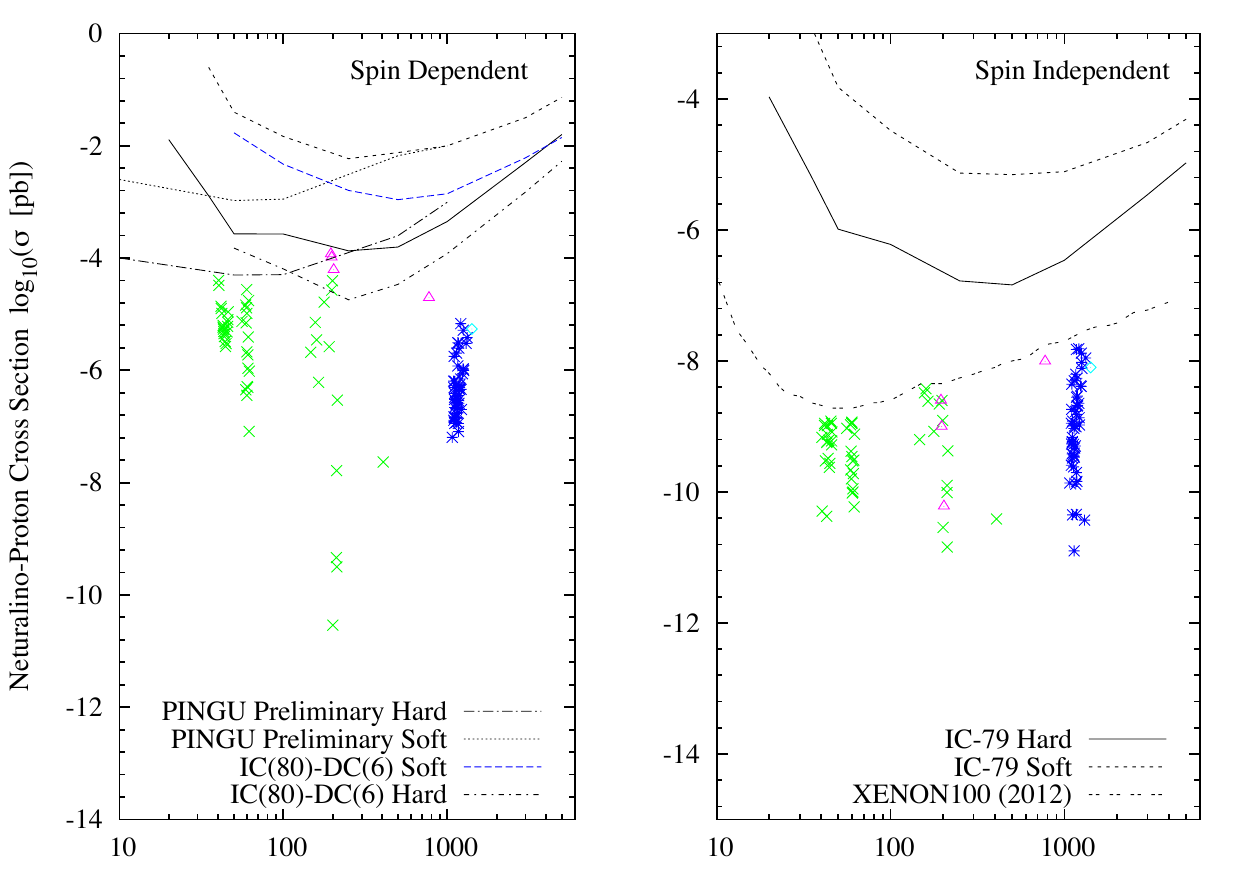} \\
    \includegraphics[scale=0.6]{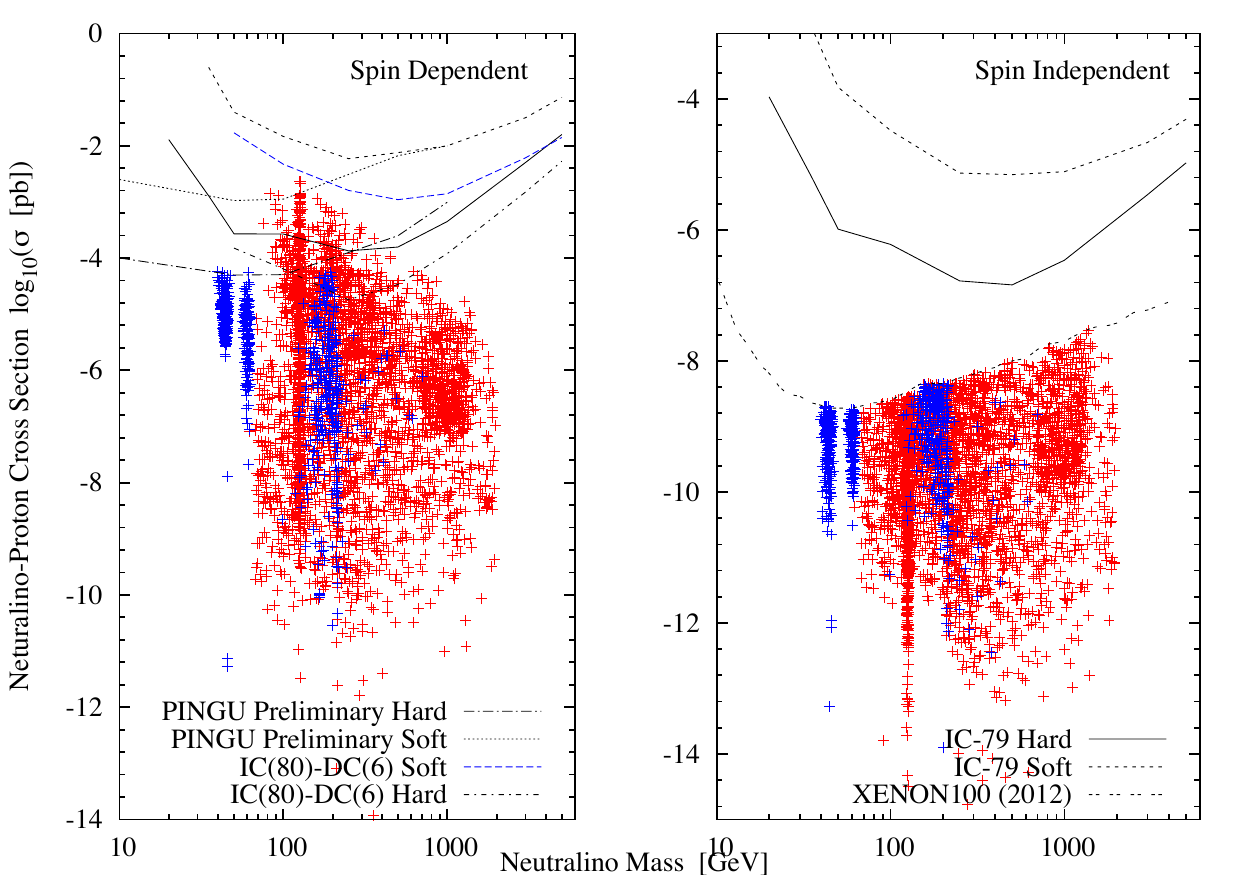} & \includegraphics[scale=0.6]{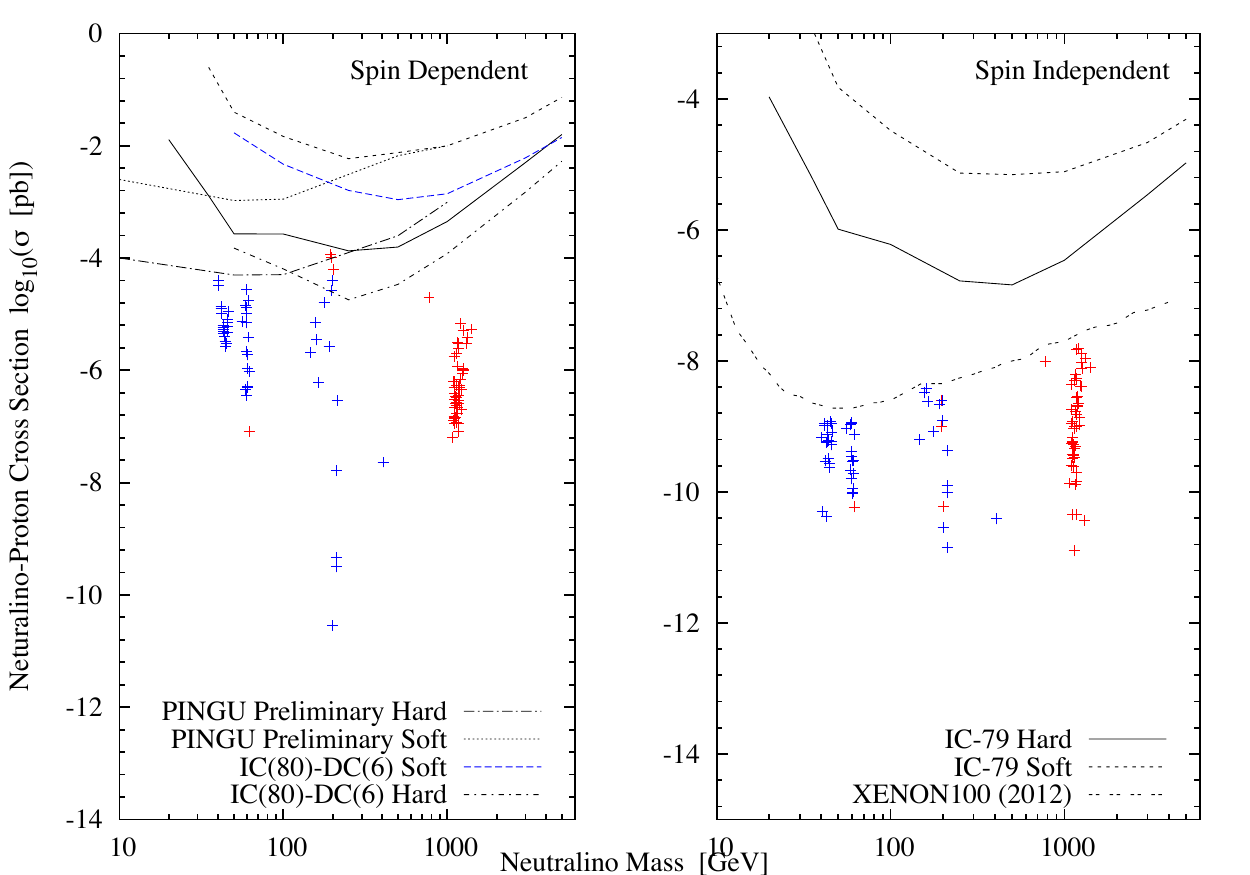}
  \end{tabular}
  \caption{Shown here are the results of the scan for neutralino-proton cross sections for spin dependent (\emph{left}) and spin independent (\emph{right}). We also indicate the limits set by IceCube-79~\cite{:2012ef} and the expected limits from PINGU~\cite{Katz}. \emph{Upper Panels}: We plot the muon flux with the color and symbol scheme of Figures~\ref{fig:DensityvMass}~and~\ref{fig:flux} to show different neutralino compositions. \emph{Lower Panels}: Same plot as the upper panel with where the color scheme is now to distinguish models that have a 10\% or better count rate increase in the neutrino flux with PINGU (see Figure~\ref{fig:flux}) (in blue) to all other models (red). The \emph{left two columns} show all models from Figure~\ref{fig:DensityvMass} while the \emph{right two columns} show only those with the correct neutralino relic density.}
  \label{fig:cross}
\end{figure*}

Figure~\ref{fig:cross} shows again the results of our MSSM parameter space scan, but highlighting this time the sensitivity of neutrino telescopes to spin-dependent (\emph{left}) and spin-independent (\emph{right}) neutralino-proton interactions. We note that, as shown for example in references \cite{1loop1,1loop2}, both spin-dependent and spin-independent neutralino-proton cross sections receive important corrections from 1-loop processes, especially if the lightest neutralino is wino- or higgsino-like. We do not include those corrections here, but the Reader should be aware of this important caveat.

In the top two panels, the color-coding highlights, as in the previous figures, the neutralino composition. In the two lower panels, instead, we indicate with blue crosses models that have a 10\% or more better count rate with the addition of the PINGU sub-array. In all panels, we indicate with various lines results from the 2012 IceCube 79-string results~\cite{:2012ef} as well as the expected limits with the proposed upgrade to PINGU~\cite{Katz}. The \emph{left two columns} show all models from Figure~\ref{fig:DensityvMass} while the \emph{right two columns} show only those with the correct neutralino relic density.

As expected from neutrino telescopes looking for a signal from the Sun, better limits compared to MSSM predictions are set on spin dependent cross sections rather than on spin independent interactions, with {\em PINGU making roughly an order of magnitude increase} over the current IceCube configuration and cutting deep into the MSSM predictions over the 50-100 GeV neutralino mass range. As expected, the impact of PINGU decreases with increasing neutralino mass, and thus neutrino energy. Neutralinos with wino-higgsino composition (which are thermally under-abundant, but whose abundance might be boosted by non-thermal mechanisms) are well within the PINGU estimated sensitivity. Many pure higgsino and bino-higgsino models are also well within the PINGU reach, for a remarkably wide range of masses.

\begin{figure*}[!h]
 \begin{tabular}{c | c}
   \includegraphics[scale=0.6]{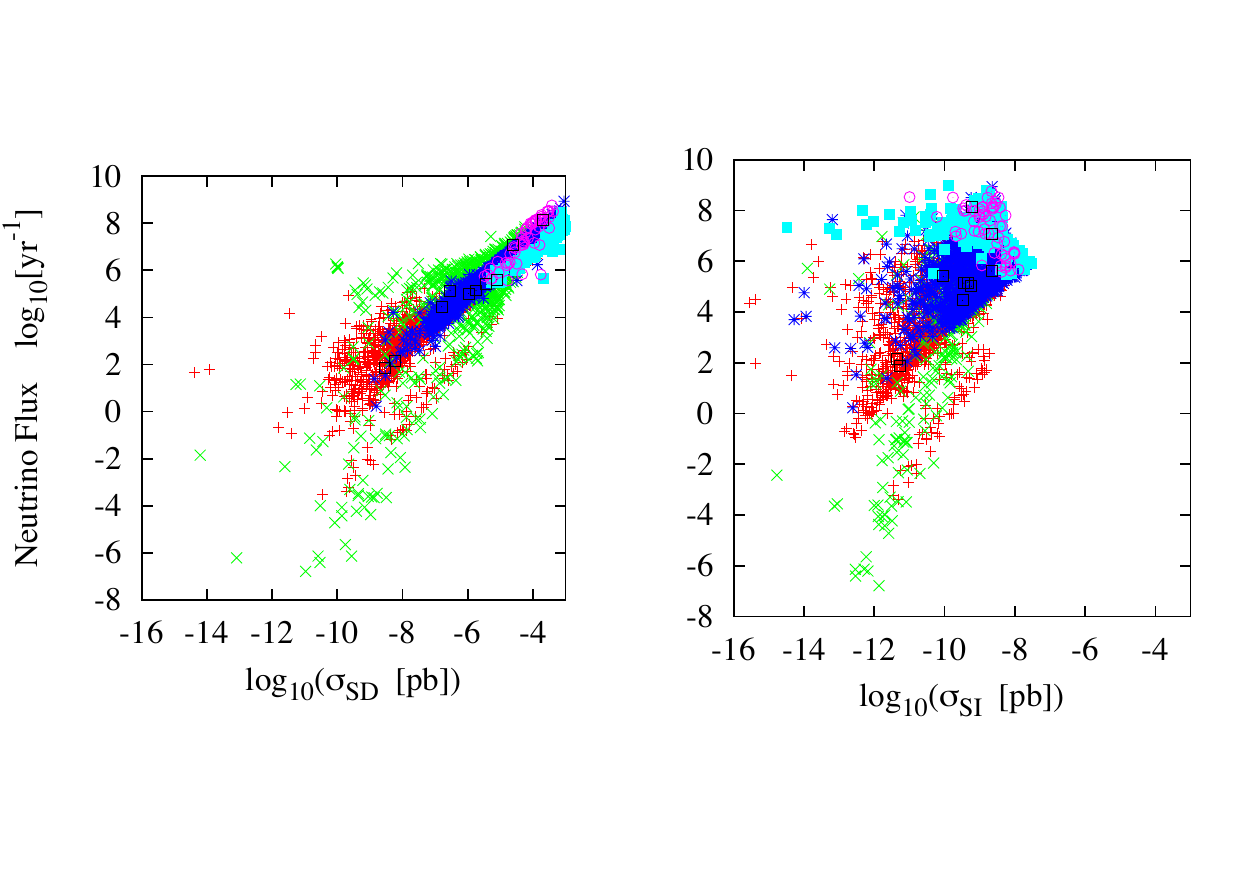} & \includegraphics[scale=0.6]{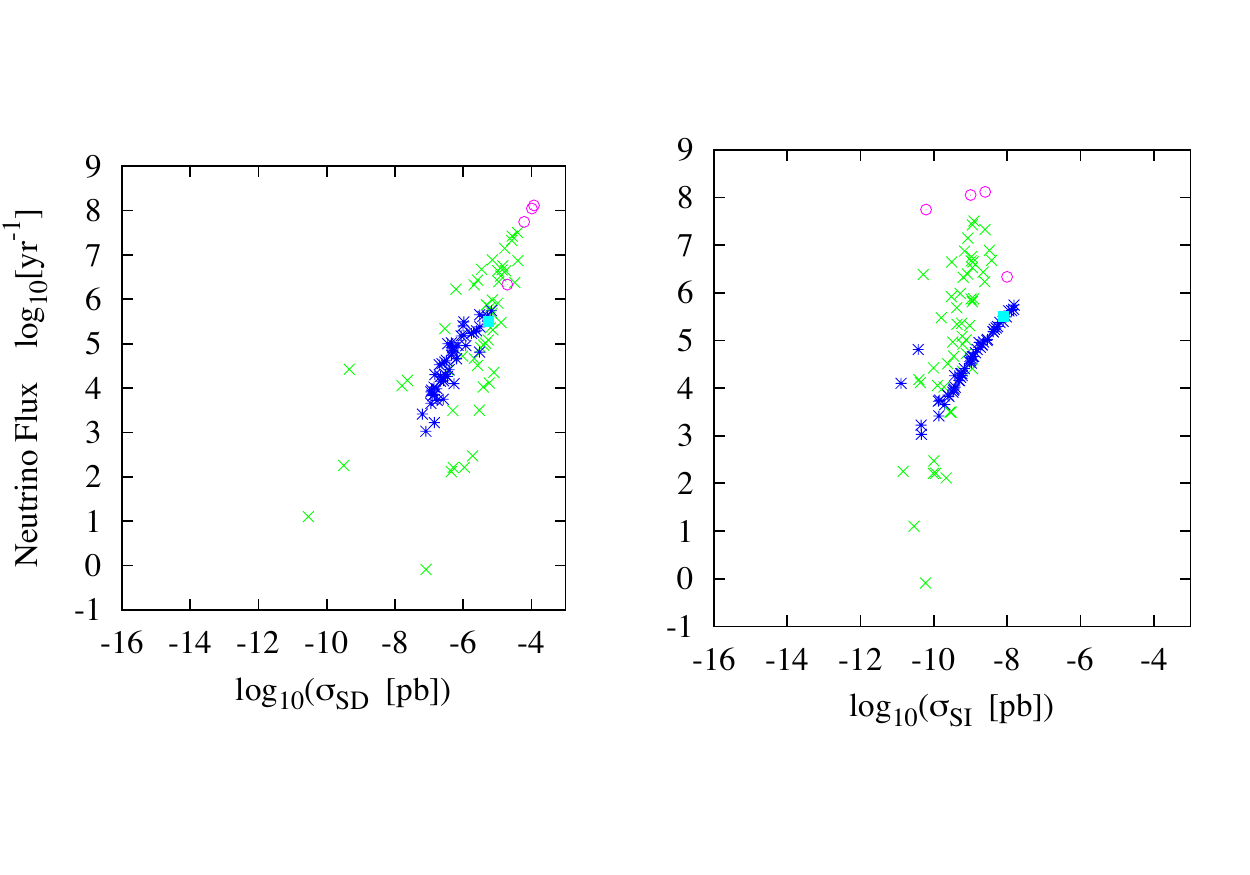} \\
   \includegraphics[scale=0.6]{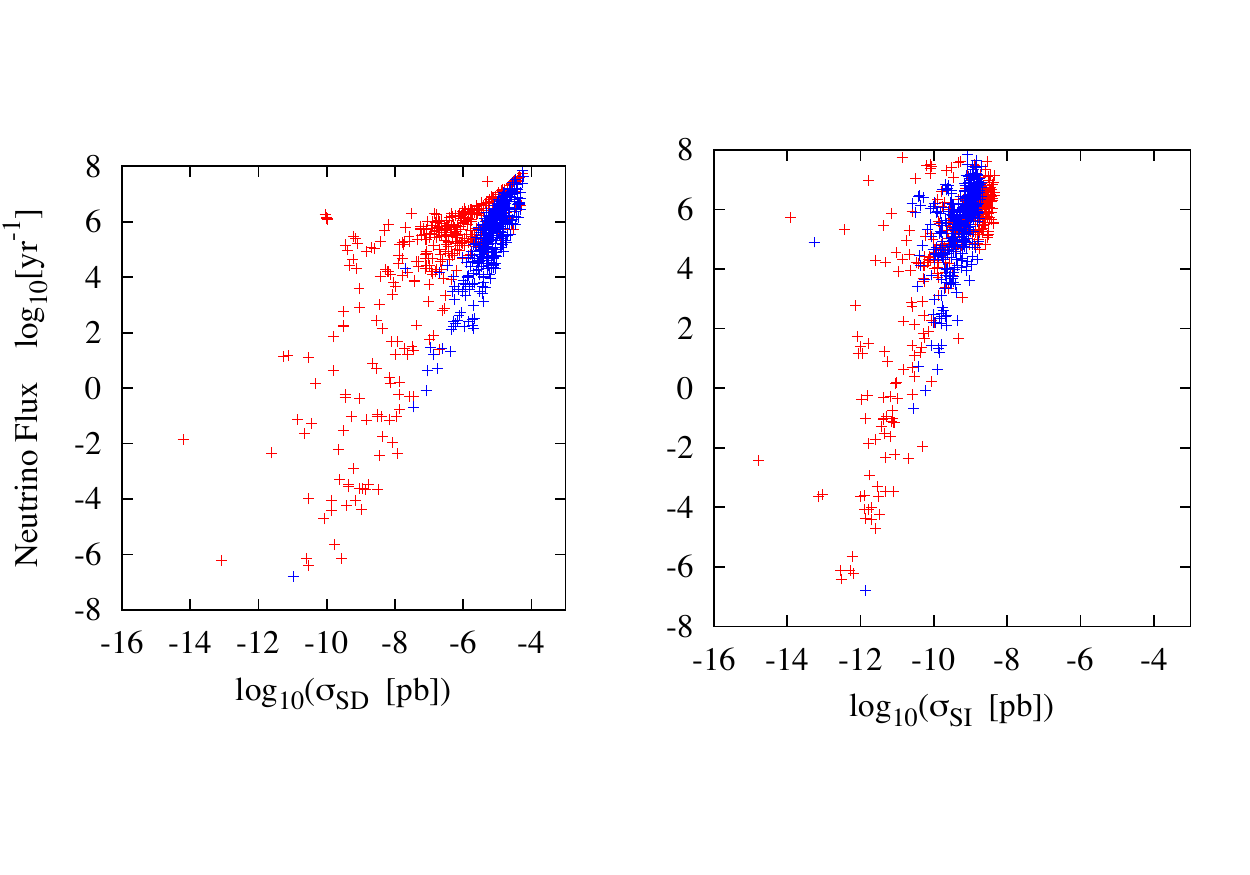} & \includegraphics[scale=0.6]{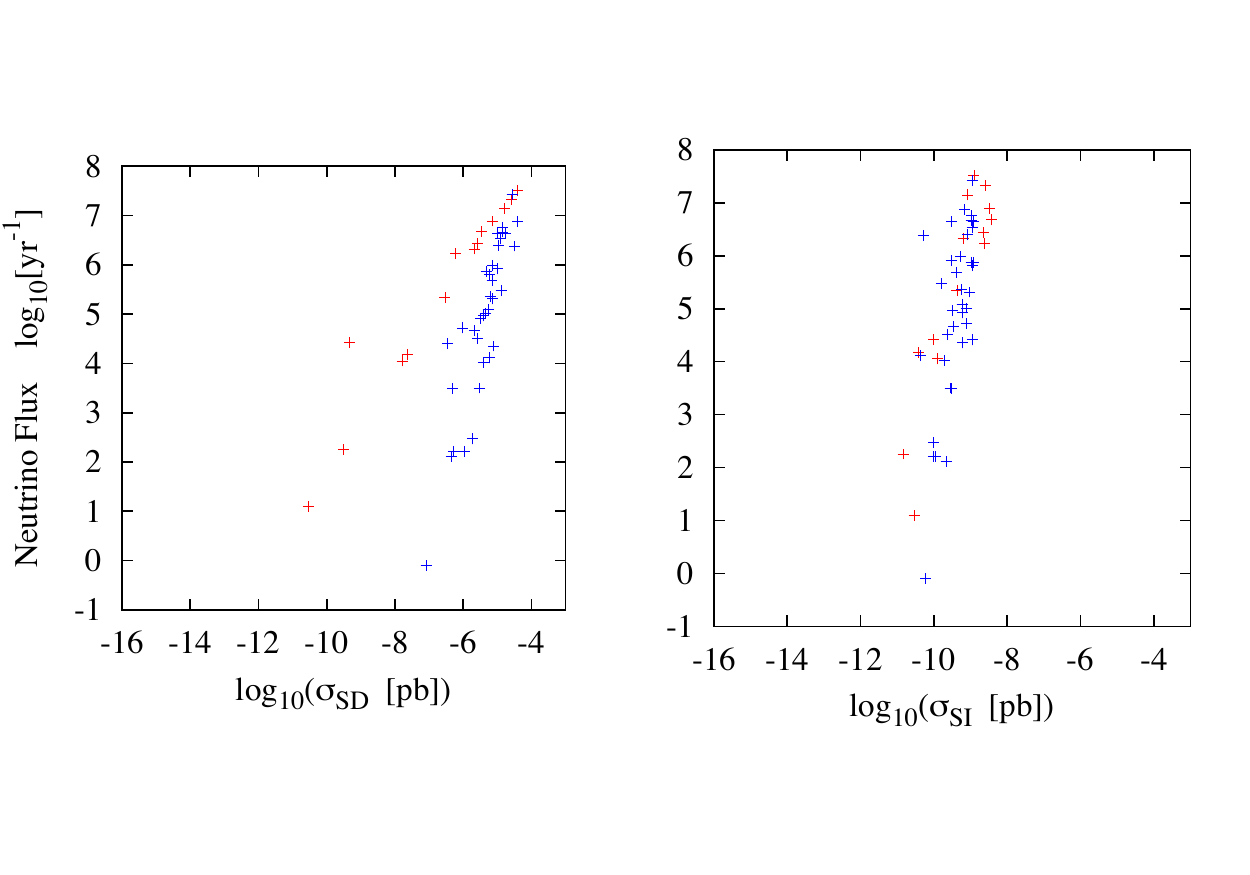}
 \end{tabular}
  \caption{Neutrino flux rate (i.e. integrated over the effective area) versus the neutralino-proton cross section, spin dependent (\emph{first and third columns}) and spin independent (\emph{second and fourth columns}). \emph{Upper Panels}: We use the same symbol and color scheme used in Figures~\ref{fig:DensityvMass}~and~\ref{fig:flux} to distinguish among different neutralino compositions. \emph{Lower Panels}: Here we highlight models in blue that have a 10\% or greater increase in the neutrino flux with PINGU (see Figure~\ref{fig:flux}). The \emph{left two columns} incorporate all models from Figure~\ref{fig:DensityvMass} while the \emph{right two columns} show only models in with relic densities in agreement with the WMAP results.}
  \label{fig:spin}
\end{figure*}

The rate of neutrinos from the center of the Sun correlates with the capture rate inside the Sun, with a close to one-to-one correspondence for models where equilibrium between capture and annihilation occurs. Figure~\ref{fig:spin} elaborates on this point, and examines the cross correlation between both spin-dependent (\emph{left}) and independent (\emph{right}) neutralino-proton cross sections and the neutrino flux from the Sun. The two upper panels show models broken down by neutralino composition, while the lower panels highlight in blue models for which PINGU makes a 10\% or greater increase in IceCube/DeepCore's detection rate. The \emph{left two columns} show all models from Figure~\ref{fig:DensityvMass} while the \emph{right two columns} show only those with the correct neutralino relic density.

The correlation between spin-dependent scattering cross section and the flux of neutrinos from the Sun is clearly visible in the left panels. It is especially strong for higgsino and mixed-higgsino states, that are almost perfectly aligned along a single line on the log-log plot. A considerable spread is present for some purely bino (green X's) and for purely  wino-like (red +'s) states: this is due to the potentially highly suppressed capture rate, that depends critically on the higgsino content. We notice, however, that for a large neutrino flux, models in the MSSM with an acceptable relic density do predict a lower limit on the spin-dependent cross section for a given neutrino flux, and vice-versa.

The spin-independent scattering cross section, instead, allows for a much looser set of conclusions. For example, models with a large neutrino flux can have a very highly suppressed spin-independent neutralino-proton cross section, and vice-versa a large scalar cross section does not guarantee a large neutrino flux.

The lower panels show that for those models where PINGU will see a significantly enhanced neutrino flux, a typically larger spin-dependent cross section is expected, for a given neutrino flux. Again, only much looser conclusions can be drawn for spin-independent interactions.

\section{Discussion and Conclusions}\label{sec:conclusions}

We assessed the role of neutrino telescope searches for the flux of high-energy neutrinos from the center of the Sun that could result from the annihilation of supersymmetric neutralinos captured and gravitationally sunk inside the Sun.  We carried out an extensive  scan of the MSSM parameter space, implementing all relevant constraints from direct collider searches for supersymmetric particles as well as indirect constraints from flavor physics and precision measurements.  We selected models with a Higgs sector compatible with the recent ATLAS and CMS results \cite{atlas, cms}, and we insisted on models having a thermal relic density within or below the observed abundance of dark matter (for under-abundant models, we assumed that non-thermal production be responsible for neutralino being the only dark matter constituent). We also detailed on models that have a thermal relic density within the observed universal dark matter density.

We found that the IceCube telescope is testing and constraining significant portions of the MSSM neutralinos across a surprisingly wide range of lightest neutralino masses, from below 100 GeV to above one TeV. Models that are excluded or that will be soon within reach of IceCube always feature a significant degree of higgsino mixing (which enhances the capture rate in the Sun). 

We studied the role of the proposed additional sub-array PINGU in the search for MSSM neutralinos, and showed that, compared to IceCube alone, PINGU brings about a 2-3 orders of magnitude improvement in the detectable neutrino flux; the figure is reduced to at most 30\% (with typical gains on the order of 5\%) for IceCube plus DeepCore. Models that are particularly well-suited to give a signal at PINGU are dominantly bino-like. We find that the recent Higgs mass measurement and of the generic SUSY searches with the LHC does not add any qualitative information to the SUSY parameter space and its relation to neutralino dark matter. What small differences these results do add to the parameter space are limited to bino-like neutralinos, and, specifically, to a mass range where the Higgs resonance is instrumental in producing the observed thermal relic density.


\begin{acknowledgments}
\noindent  Stefano Profumo is partly supported by the US Department of Energy under Contract DE-FG02-04ER41268.
\end{acknowledgments}

\end{document}